# ALGORITHMIC DERIVATION OF HUMAN SPATIAL NAVIGATION INDICES FROM EYE MOVEMENT DATA

Sobhan Teymouri[1]    Fatemeh Alizadehziri[1]    Mobina Zibandehpoor[1]
Mehdi Delrobaei[1,*]

[1]Mechatronics Department, Faculty of Electrical Engineering, K. N. Toosi University of Technology, Tehran, Iran
[*]Corresponding author: delrobaei@kntu.ac.ir

## ABSTRACT

Spatial navigation is a complex cognitive function involving sensory inputs, such as visual, auditory, and proprioceptive information, to understand and move within space. This ability allows humans to create mental maps, navigate through environments, and process directional cues, crucial for exploring new places and finding one's way in unfamiliar surroundings. This study takes an algorithmic approach to extract indices relevant to human spatial navigation using eye movement data. Leveraging electrooculography signals, we analyzed statistical features and applied feature engineering techniques to study eye movements during navigation tasks. The proposed work combines signal processing and machine learning approaches to develop indices for navigation and orientation, spatial anxiety, landmark recognition, path survey, and path route. The analysis yielded five subscore indices with notable accuracy. Among these, the navigation and orientation subscore achieved an $R^2$ score of 0.72, while the landmark recognition subscore attained an $R^2$ score of 0.50. Additionally, statistical features highly correlated with eye movement metrics, including blinks, saccades, and fixations, were identified. The findings of this study can lead to more cognitive assessments and enable early detection of spatial navigation impairments, particularly among individuals at risk of cognitive decline.

## 1 Introduction

The human ability to navigate through familiar environments, such as one's residence, even under low light conditions, is underpinned by a sophisticated cognitive mechanism referred to as spatial navigation Chen et al. [2023a], Wilkins [2011]. Humans use spatial navigation as a complex cognitive process that is important in finding their way around the environment by utilizing different senses and areas of the brain McNamara and Chen [2022], Chen et al. [2023b], Garg et al. [2024], Verghese and Blumen [2022]. It involves cues such as landmarks and information on self-motion to determine positions and achieve goals Roth et al. [2020].

A thorough understanding of spatial navigation is essential for improving destination efficiency and reducing anxiety in unfamiliar settings. Assessing spatial navigation is crucial for evaluating cognitive health, especially in neurological and neurodegenerative diseases Roth et al. [2020]. Spatial navigation tasks can detect structural changes in subcortical brain areas related to cognitive decline risk Chen et al. [2023b]. Different neurodegenerative conditions see impaired spatial navigation as a symptom at the onset; thus, it can be a valuable predictor of dementia in subjective cognitive decline patients or those with mild cognitive impairment Tangen et al. [2022]. These deficits worsen with aging, highlighting the urgent need for efficient assessment tools such as the Virtual Environments Navigation Assessment (VIENNA), which evaluates spatial navigation abilities Rekers and Finke [2024a]. This research is critical for detecting cognitive impairments and guiding clinical decisions.

Spatial navigation abilities have been evaluated with various unique instruments, and different interpretations have been provided, such as physical or behavioral tasks that are assessed with various tools Hamre et al. [2020], Castilla et al. [2022]. However, some limitations of these tasks can influence their validity. Some complex experimental paradigms may be too challenging for cognitively impaired people, while others may have excessive instructions that hinder their participation, especially in clinical settings. Also, conventional assessments often fail to reproduce real-world navigation challenges and lack ecological validity Rekers and Finke [2019]. Other tools, such as virtual environments, can test validity and feasibility among healthy subjects. Still, cybersickness poses a significant challenge, affecting performance and data reliability Belmonti et al. [2015], Newman and McNamara [2021], Pai [2023].

Additionally, motor skills can be mistaken for cognitive performance during active navigation, complicating the assessment of spatial abilities Rekers and Finke [2024b]. Advanced techniques such as electroencephalogram (EEG) assessments provide insights into navigational learning and cognitive load discrepancies between experts and novices by monitoring brain activity. Still, the discomfort caused by the equipment can be distracting and interfere with accurate results, making them inefficient for assessing spatial navigation Miyakoshi et al. [2021], Keskin et al. [2019], Postelnicu et al. [2012], Hanzal et al. [2023]. Game-based methods evaluate navigation and visuospatial abilities, addressing the interplay between these tasks Garg et al. [2024].

This study presents new indices that use automation and algorithms to evaluate human navigation abilities. These indices utilize computational algorithms, machine learning, and data analytics to objectively measure spatial navigation skills such as navigation and orientation, route knowledge, and spatial anxiety. The goal is to create better assessment tools for clinical and research purposes, improving diagnostic methods. Eye movements will be recorded using electrooculogram (EOG) while participants watch a virtual reality-based video.

This paper is organized as follows. Section II surveys the related works. Section III proposes the materials and methods of the work. The results and discussions are elaborated in Section IV. Finally, Section V presents the conclusion and remarks for future research.

## 2 Related works

Numerous tools have been developed to assess human navigation abilities. This section reviews studies and methods for evaluating spatial navigation, highlighting existing techniques, their applications, and insights into spatial cognition across diverse populations. Many physical tasks have been designed to assess spatial navigation. Charlotta Hamre *et al.* utilized the Floor Maze Test (FMT) in individuals with dementia or Alzheimer's disease Hamre et al. [2020], while Riccardo Maria Martoni *et al.* used the Walking Corsi Test across all age groups to measure subjects' performances on space representation tasks Castilla et al. [2022]. However, the FMT may not fully capture the spatial abilities of cognitively healthy individuals, and the Walking Corsi Test is criticized for lacking ecological validity in real-world navigation scenarios.

Virtual-reality (VR) tasks offer several advantages over physical activities in assessing spatial navigation. The VR provides a controlled and ecologically valid environment for studying spatial navigation Creem-Regehr et al. [2022]. Kiran Ijaz *et al.* used a VR platform to perform a landmark recall test. Participants identified landmarks within a virtual environment, with the assessment focusing on correct recalls, navigational mistakes, task duration, engagement levels, and stress levels Ijaz et al. [2019].

Tanya Garg *et al.* tested spatial abilities using the mobile game Sea Hero Quest (SHQ), which included wayfinding, path integration, and spatial memory tasks. Participants also completed the mental rotation, visuospatial processing, and working memory tests, revealing a positive relationship between performance in the wayfinding task and visuospatial skills, indicating that different navigation aspects require distinct skill sets Garg et al. [2024]. Alan J. Smith *et al.* used a virtual game where participants navigated a virtual city to locate landmarks and later recalled and delivered items to these locations, assessing spatial navigation and memory recall abilities based on the time taken to find each location Smith et al. [2024]. However, assessing spatial navigation in VR without questionnaires has limitations, as questionnaires provide essential insights into participants' experiences, cognitive strategies, and challenges, which are crucial for interpreting performance data, understanding participant engagement, and comparing results across studies.

Several questionnaires evaluate spatial navigation and anxiety. The Navigation Strategy Questionnaire (NSQ) assesses individual differences across three subscales: Egocentric Spatial Updating, Survey-Based Strategy, and Procedural Strategy Garg et al. [2024]. The Santa Barbara Sense of Direction (SBSOD) Scale is a 15-item Likert scale measuring the overall sense of direction Hegarty et al. [2002]. The Wayfinding Questionnaire (WQ), with 22 items, evaluates navigation, orientation, spatial anxiety, and distance estimation De Rooij et al. [2019a]. Van der Ham *et al.* recommended three steps: first, directly inquire about potential navigation problems; second, use the WQ to assess patients' perception of their navigational skills and spatial anxieties; lastly, employ the Leiden Navigation Test (LNT) to measure navigation abilities objectively, confirming or clarifying reported deficits. Normative data for WQ and LNT are provided from a large population sample De Rooij et al. [2019b].

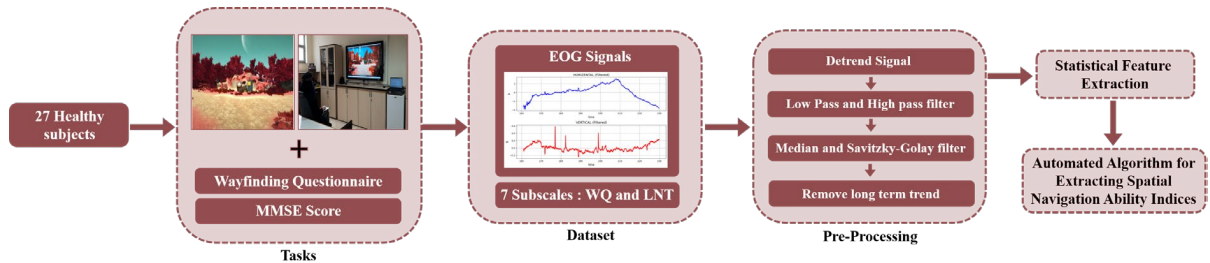

Figure 1: Architecture of the proposed model.

Although VR environments address the limitations of physical tasks, they can be enhanced with additional inputs like EEG, EOG, or motion capture data for deeper insights into neural activity, eye movements, and body dynamics during navigation tasks. Chin-Teng Lin *et al.* assessed spatial navigation using a VR-based tunnel driving environment where participants tracked their 3D position relative to the starting point while viewing animations of passages through a 3D virtual tunnel. At the end of each passage, they indicated the homing direction by pressing buttons, reflecting their navigation strategy and reference frame usage. Participants were categorized into allocentric or egocentric groups based on their performance, and EEG signals were analyzed to compare brain activity between these navigation strategy groups Lin et al. [2009].

Makoto Miyakoshi *et al.* used EEG synchronized with motion capture in a virtual maze to provide behavioral evidence for navigational learning. Participants navigated mazes multiple times in the audio maze study to assess spatial learning, with EEG and motion capture data collected to analyze brain dynamics and movement patterns associated with navigation Miyakoshi et al. [2021]. While EEG effectively records brain activity, it does not capture eye movements, which are crucial for assessing spatial navigation as they reflect cognitive processes involved in planning and decision-making Zhu et al. [2022]. Additionally, motion capture clothing can affect joint angle estimation due to its interaction with underlying bones Lorenz et al. [2022].

Thus, eye movements should be recorded using eye tracking or EOG. Merve Keskin *et al.* combined EEG and eye tracking to study cognitive processes in expert and novice map users. Participants first studied a map without time limits and then drew a sketch, comparing spatial information processing between groups. A second experiment involved more stimuli to examine the effect of task difficulty on cognitive processes, with spatial navigation assessed through EEG and eye-tracking data Keskin et al. [2019]. The combined use of EEG and EOG provides a clearer picture of cognitive processes in spatial navigation. However, this combination has only been tested in physical tasks. Negar Alinaghi *et al.* investigated gaze behavior and visual attention during wayfinding, identifying distinct patterns across self-localization, route planning, monitoring, and goal recognition stages. Their findings showed that spatial familiarity significantly influences navigation strategies, supported by eye-tracking data Alinaghi and Giannopoulos [2024].

Looking at the reviewed articles, many tasks focus only on one spatial navigation subscale, such as spatial memory. Our research employs a VR-based task that records participants' eye movements via EOGs to assess spatial navigation across five scores, each analyzed within subscales. We selected the WQ for its multiple sub-scores, which offer additional insights into navigation strategies. By using non-invasive EOG, we address the limitations of past studies, enabling a more detailed and analysis of eye movements in relation to spatial navigation performance. This approach enhances the understanding of cognitive processes involved in navigation.

## 3 Materials and Methods

Our approach to developing spatial navigation indices includes four essential stages: study design and data collection, pre-processing of the data, feature extraction, and index generation (Fig. 1).

### 3.1 Dataset Description

Twenty-seven healthy university students, averaging 21.81 years (SD=1.96), participated in the study. Informed consent was obtained from all participants, and the study followed the Declaration of Helsinki principles. Before beginning the tasks, each participant underwent a cognitive assessment using the mini-mental state examination (MMSE) test

Tombaugh and McIntyre [1992] to ensure normal cognitive function. The participants then completed the WQ, which included three main subscales: Navigation and Orientation, Spatial Anxiety, and Distance Estimation. They were then seated in front of a monitor to perform the LNT, during which their eye movements were recorded using an EOG headband Zibandehpoor et al. [2024].

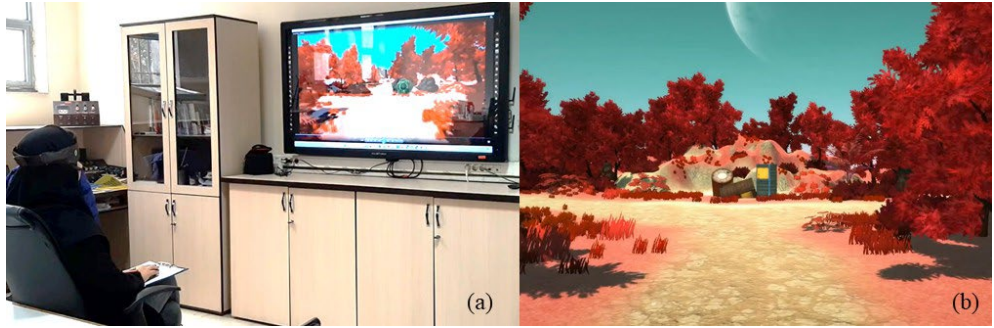

Figure 2: The experiment materials: (a) The participant is undergoing the assessment. (b) A sample of the virtual environment.

A commercial two-channel wireless EOG headband (Zehnafzar Rayan Co., Isfahan, Iran) with five electrodes was used to record eye movement signals (Fig. 2(a)). The vertical and horizontal eye gaze patterns were recorded. A sample pattern can be seen in Fig. 3. Data was collected at a sampling rate of 250 Hz, and the recorded signals were then transmitted to a PC via Bluetooth for real-time display and monitoring.

### 3.2 Pre-processing

First, the linear trend of the signal was removed. Next, low-pass and high-pass filters were applied to eliminate noise. A low-pass filter with a cutoff frequency of 20 Hz was used to remove high-frequency noise in the EOG signal, while a high-pass filter with a cutoff frequency of 0.05 Hz was implemented to neutralize slow noise. A median and Savitzky-Golay (SG) filters were employed to reduce noise and achieve a smoother signal. The SG filter was specifically selected for its superior noise reduction performance and ability to preserve essential data features Hasan et al. [2022].

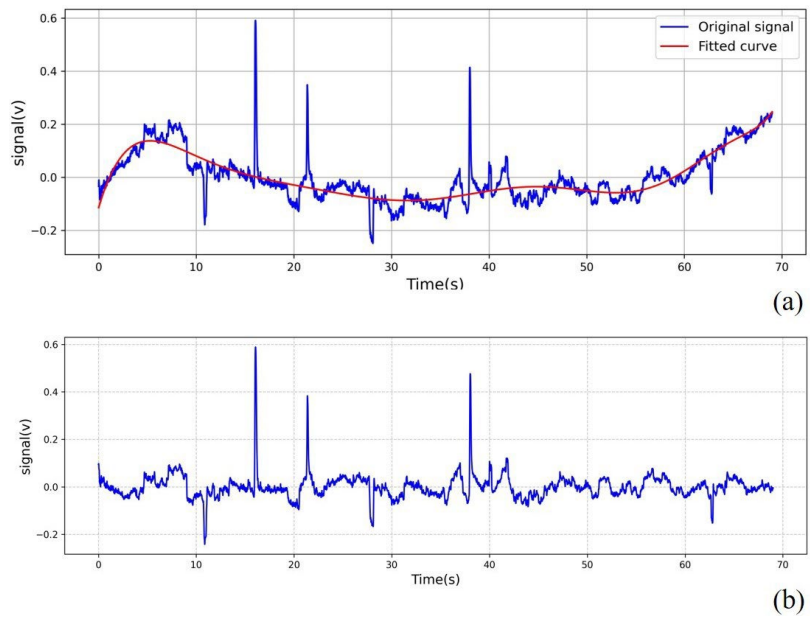

Figure 3: Signal processing procedure (a) A sample of the filtering process. (b) A sample of pre-processing stage output.

After applying the SG filter, a marked improvement in signal quality was visually observed, resulting in more robust and reliable analysis outcomes. Based on these observations, a window length of seven points and a polynomial order of two were selected for the SG filter. In the final stage, the overall trend of the signal was removed. The signals were fitted to a polynomial to address long-term baseline drift, with the degree determined individually for each subject using the Bayesian Information Criterion (BIC) method Baray et al. [2023]. The fitted values were then subtracted from the original signals. Fig. 3(a) demonstrates the line-fitting process applied to the signal. At the same time, Fig. 3(b) presents a sample of the signal after undergoing the pre-processing steps, along with the resulting output.

### 3.3 Statistical Feature Extraction

After signal pre-processing, the subsequent step involved feature extraction, focusing on identifying and analyzing key attributes that characterize the signals for each participant. This step is vital for capturing the underlying patterns and behaviors within the signals, offering valuable insights into the data. In this study, 22 statistical features were selected to extract meaningful information from both vertical and horizontal signals. These features encompass a broad spectrum of statistical measures, each providing a distinct perspective on the signal data. Table 1 summarizes these features, presenting their mathematical definitions and interpretations.

Each feature was meticulously chosen to enhance the understanding of the signal characteristics, enabling a robust analysis suitable for further modeling and data interpretation. These features capture essential aspects of the signal, including central tendency, dispersion, distribution shape, and energy dynamics. For example, the mean, median, and mode offer insights into the signal's central tendency, while variance, standard deviation, and interquartile range describe its dispersion. Higher-order statistics, including skewness and kurtosis, provide information about the asymmetry and peakedness of the signal distribution. Moreover, features such as the root mean square and signal magnitude area quantify the signal's overall energy. In contrast, spectral features such as the spectral centroid and bandwidth characterize its frequency domain properties. Advanced features, including lyapunov exponents and detrended fluctuation analysis, are utilized to explore the signal's complexity and dynamic behavior.

### 3.4 Spatial Navigation Indices

An iterative algorithm was developed to assess spatial navigation ability by integrating machine learning techniques with statistical analysis. The primary objective was to identify the optimal combination of features from a predefined set that most accurately predicts the target variable while minimizing the risk of overfitting through regularization, as detailed in Algorithm 1.

A linear regression model within a cross-validation framework was employed to assess the predictive performance of various feature combinations systematically. The selection of features was based on their potential relevance to the subscores. A K-fold cross-validation approach with five splits was utilized to ensure a robust model performance evaluation. This method mitigates the risk of overfitting by averaging the model's performance across multiple training and testing phases. The linear regression model was trained on a selected feature pair for each fold, and its predictive accuracy was evaluated on the corresponding test set.

All possible pairs of features were generated and subjected to cross-validation to identify the optimal feature pair. The mean squared error (MSE) was used as the evaluation metric, quantifying the average squared difference between observed and predicted values. A regularization term ($\lambda = 0.005$) was added to the MSE to refine the model selection. This regularization penalizes large coefficients in the regression model, thereby preventing overfitting and promoting simpler, more generalizable models. The final score for each feature combination was calculated as the sum of the average MSE and the regularization penalty. The combination with the lowest score was deemed the optimal choice.

The selected feature pair was then combined into a new feature, and the process was repeated. This iterative process continued, adding each new feature combination until no further significant improvement in model accuracy, as defined by a predefined threshold of 0.01, was detected. This approach progressively refined the model and enhanced its accuracy in predicting subscores. The analysis identified the best combination of features, and their corresponding coefficients provided insights into the linear relationship between these features and the target variable, facilitating further interpretation and application in the assessment of spatial navigation.

### 3.5 Evaluation Metrics

The proposed method was evaluated by analyzing the correlation of statistical features with blink, saccade, and fixation events. Given that the normality assumption was violated for several features, as determined by the Shapiro-Wilk test ($p \leq 0.05$), Spearman correlation coefficients were employed for the correlation analysis. This method is more suitable for non-normally distributed data and captures monotonic relationships between variables.

**Algorithm 1** Calculate index for estimating each subscale.
1: Define the number of folds for cross-validation ($k = 5$)
2: Define a regularization parameter ($\lambda_{reg} = 0.005$)
3: Get the target variable (y = dataframe['subscale_name'])
4: Initialize variables to track the optimal best_score, best_combination, and best_coefficients
5: Identify the columns to be used as features (selected_columns)
6: Generate all possible pairs of features from selected_columns (feature_combinations)
7: **for** each feature combination in feature_combinations **do**
8:    Extract the corresponding features (X) from the data frame based on the combination
9:    Initialize a list to store MSE for each fold (mse_list)
10:   Perform k-fold cross-validation:
11:   **for** each fold **do**
12:      Split data into training and testing sets
13:      Train a linear regression model on the training set
14:      Predict the target variable on the testing set
15:      Compute the MSE for the predictions
16:      Append the MSE to mse_list
17:   **end for**
18:   Calculate the average MSE from mse_list
19:   Compute the regularization penalty by the model's coefficients:
20:   (penalty = $\lambda_{reg} \times$ sum of the absolute values of the model's coefficients)
21:   Calculate the score by adding the average MSE and the penalty
22:   **if** the score is lower than best_score **then**
23:      Update best_score with the new score
24:      Update best_combination with the current feature combination
25:      Update best_coefficients with model's coefficients
26:   **end if**
27: **end for**
28: **Output:** the best feature combination, best score, and best coefficients
29: Save the best feature combination with their best coefficients as a new feature
30: **Repeat** lines 1-29 iteratively until the improvement in MSE (new – previous) falls below a predefined threshold

To assess the validity of the generated indices, various evaluation metrics were employed, including MSE, mean absolute error (MAE), root mean squared error (RMSE), mean average percentage error (MAPE), and the R² score. The MAE measures the average magnitude of errors in the predictions without regard to their direction. MSE represented the average squared differences between predicted and actual values, assigning greater weight to larger errors. RMSE, as the square root of MSE, provided an error metric on the same scale as the original data. MAPE quantified the accuracy of predictions as a percentage, measuring the average absolute percent error between predicted and actual values. The R² score, or the coefficient of determination, indicated the proportion of variance in the dependent variable that could be explained by the independent variables, with values closer to one suggesting a better model fit.

## 4 Results and Discussions

Employing the proposed algorithm, a set of formulas was derived for each score by combining statistical features through addition or subtraction, resulting in dimensionless numerical values. This standardization simplifies the interpretation of the results and makes further analyses more straightforward. Dimensionless scores are also more versatile and robust, as they are less sensitive to measurement errors and inconsistencies. The coefficients in these formulas ensure the final values are dimensionless, enabling accurate estimation of each subscore. These dimensionless values, along with fitted line data based on the dataset, were utilized to estimate the subscores. The performance of the extracted indices is presented in Table 2. The navigation and orientation score demonstrated the best performance among the various scores, while the location allocentric score exhibited the least accurate estimations.

Upon reviewing the model's performance, the proposed algorithm successfully estimated 5 out of the seven subscores with reasonable accuracy.

The navigation and orientation (NO) score was derived as

$$NO = (-1.03) \times (8000\,SB_v + 2\,TK_h - 0.14\,ZCR_h - 9\,SK_h$$
$$+ 8\,MO_h - 0.02\,ER_v + 10\,MA_v - 30\,RMS_v) + 110.64. \quad (1)$$

In this and all of the following expressions, the abbreviations of the included features have been borrowed based on Table 1, and the subscripts indicate either horizontal (h) or vertical (v) signals.

The spatial anxiety (SA) score was denoted as

$$SA = (-1.13) \times (-3.6\,TK_v - 52\,MO_v + 0.9\,EN_v - 0.02\,ER_v$$
$$- 13\,TK_h - 0.02\,ZCR_v - 31\,DFA_v) - 1.84. \quad (2)$$

The landmark recognition (LR) score was extracted as

$$LR = (-1.08) \times (-0.2\,EN_v - 1.4\,RA_v - TK_h - 11\,DFA_h$$
$$- MI_h - 26\,MD_v) - 18.20. \quad (3)$$

The path survey (PS) score was formulated as

$$PS = (-0.85) \times (1.2\,SK_h - 0.4\,EN_h - 4.3\,AUC_v$$
$$+ 0.16\,KU_h) - 5.22. \quad (4)$$

The path routh (PR) score was determined as

$$PR = (-1.04) \times (0.34\,SK_v + 0.001\,SMA_v + 0.71\,TK_h$$
$$- 0.06\,EN_v - 0.006\,ZCR_v - 1.6\,DFA_h) - 0.48. \quad (5)$$

The performance results of Equations 1 and 2 for individual cases are illustrated in Fig. 4, while the results of Equations 3 and 5, which pertain to subscores with limited discrete values, are displayed as bar charts in Fig. 5. Based on the data presented, Participant 18 demonstrated exceptionally high performance in the navigation and orientation subscale while also exhibiting low levels of spatial anxiety. In contrast, participant #24, despite scoring highly on the landmark recognition subscale, performed poorly in navigation and orientation. The primary reason for this poor performance appears to be the participant's exceptionally high level of spatial anxiety, the highest among all participants in the experiment. This anxiety likely contributed to their reduced navigation performance. Additionally, among Participants 13, 6, and 17, who all scored perfectly in both landmark recognition and path route subscales, those with lower spatial anxiety (Participants 6 and 13) achieved higher scores in navigation and orientation. These findings underscore the significant role that spatial anxiety plays in influencing individuals' spatial navigation performance.

In evaluating spatial navigation, eye movement features have proven to be invaluable, offering profound insights into the cognitive mechanisms driving navigation tasks. For example, gaze fixation patterns have been instrumental in understanding how individuals plan and execute movements, particularly in complex environments where effective path selection and obstacle avoidance are crucial Hunt et al. [2019]. These fixation patterns reveal that individuals tend to focus on their destination prior to initiating movement, and subsequently shift their gaze to monitor the travel path and potential obstacles. This continuous updating of visual information is essential for successful navigation Patla et al. [2007]. Moreover, the ability to separate covert attention from gaze direction has been shown to enhance the efficiency of visual information processing. This separation allows for simultaneous monitoring of multiple locations, which is critical in dynamic navigation tasks Hodgson et al. [2019]. Evidence further suggests that specific eye movements, such as saccades and blinks, are closely tied to cognitive processes involved in spatial navigation. For instance, landmark-based navigation instructions have been associated with increased blink-related brain potentials, which in turn correlate with better spatial knowledge acquisition Wunderlich and Gramann [2021].

Studies involving virtual environments have highlighted the role of sequential eye movements, or prospection, in both planning and executing navigation tasks. This underscores a direct relationship between eye movement patterns and navigational strategies Zhu et al. [2022]. The findings highlight the importance of eye movements in visually informed planning and suggest their potential in assessing navigational abilities. Our proposed method was evaluated by examining the correlation between statistical features and key eye movement events, including blinks, saccades, and fixations.

The significance of each feature in evaluating the navigation subscores is depicted in Fig. 6. The most influential statistical features: Teager-Kaiser energy operator, skewness, and entropy were utilized in four of the derived equations.

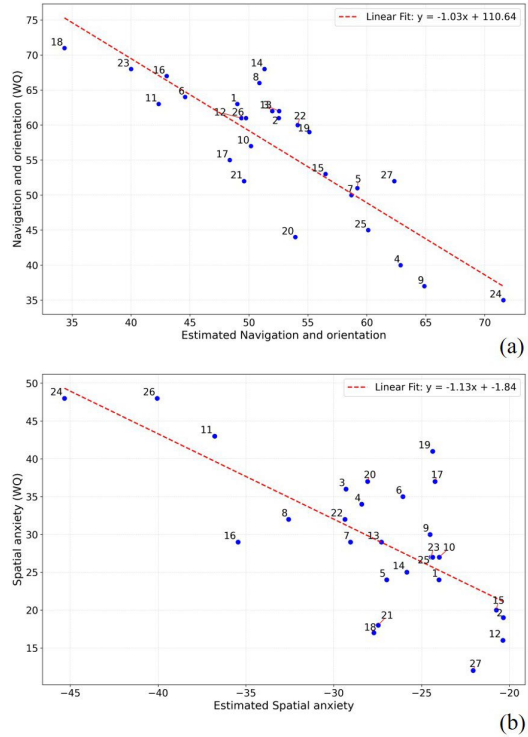

Figure 4: Scatter plots of actual versus estimated subscores for (a) Navigation and orientation, and (b) Spatial anxiety, including participant IDs annotated.

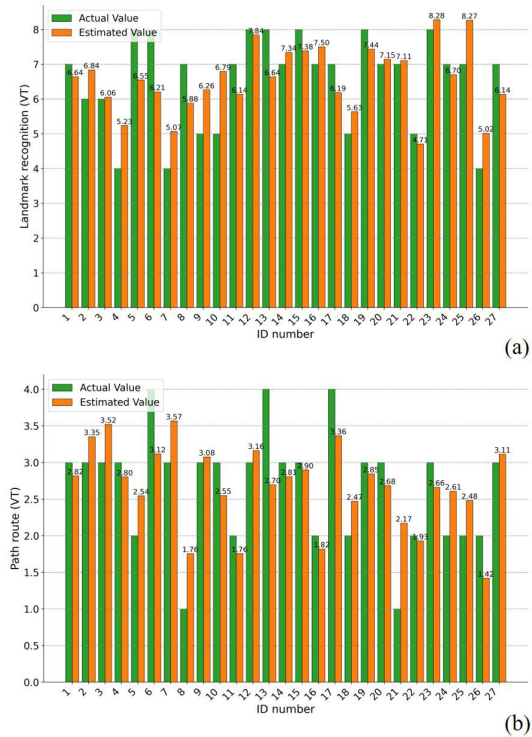

Figure 5: Bar plots of actual versus estimated subscores for (a) Landmark recognition, and (b) Path routh, with participant IDs annotated.

To understand the rationale behind selecting these features, the correlation between eye movement patterns and the extracted statistical features is shown in Table 3. Notably, the features that played a critical role in the equations also had the highest correlation with eye movement patterns. The similarity between the results in Table 3 and Fig. 6 further validates the method used. In Table 3, features marked with an asterisk (*) exhibited high correlation with only one pattern. Each derived formula incorporates different features, each representing a distinct eye movement pattern, leading to a combination of multiple eye movement patterns in the final formulas.

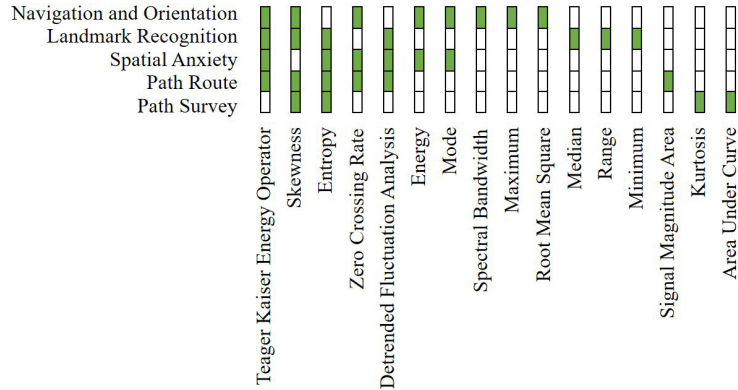

Figure 6: Bar plot of feature importance for navigation subscores.

## 5 Conclousions

This study presented a novel method for assessing spatial navigation abilities using EOG-based indices, which effectively capture various facets of navigation performance. By integrating signal processing, feature extraction, and machine learning techniques, we developed a framework to evaluate navigation-related cognitive functions. Our approach utilizes detailed eye movement data to gain insights into cognitive processes during VR-based navigational tasks. The performance of the proposed EOG-based indices was evaluated using several metrics, including $R^2$ score, MAE, MSE, MAPE, and RMSE, demonstrating the accuracy of our algorithm in estimating navigation subscores. We derived equations to estimate these subscores by combining statistical features identified as the most influential in evaluating navigation performance. The strong correlation between these features and eye movement patterns further supports the robustness of our approach. However, this study is limited by the homogeneity of the sample, which consisted of healthy university students, all within a similar age group. Future research could focus on refining EOG-based indices by incorporating additional physiological and behavioral data and video analysis to create a more accurate spatial navigation model. Including diverse populations with varying cognitive abilities and cultural backgrounds will enhance the generalizability of our findings.

Wei Zeng, Shiek Abdullah Ismail, and Evangelos Pappas. The impact of feature extraction and selection for the classification of gait patterns between acl deficient and intact knees based on different classification models. *EURASIP Journal on Advances in Signal Processing*, 2021(1):95, 2021.

## 6 Competing Interests

The authors declare no competing interests.

## 7 Tables

Table 1: Statistical Features, Definitions, and Symbols

| Variable | Definition | Symbol |
| --- | --- | --- |
| Mean | The average signal amplitude over a period Triadi et al. [2021]. | ME |
| Median | The midpoint amplitude value of a signal when sorted in order Shashidhar et al. [2023]. | MD |
| Mode | The most frequently occurring amplitude value in a signal Shashidhar et al. [2023]. | MO |
| Variance | The measure of the signal's amplitude dispersion around its mean Shashidhar et al. [2023]. | VA |
| Standard Deviation | The square root of the variance Korda et al. [2015]. | SD |
| Skewness | The measure of the signal's amplitude distribution asymmetry around the mean Triadi et al. [2021]. | SK |
| Kurtosis | A measure of the "tailedness" of the signal's amplitude distribution relative to that of a normal distribution Shashidhar et al. [2023]. | KU |
| Minimum, Maximum | The smallest and largest values in a signal Shashidhar et al. [2023]. | MI,MA |
| Range | The difference between the signal's maximum and minimum amplitude values. | RA |
| Interquartile Range | The difference between the 75th and 25th percentile amplitude values Shashidhar et al. [2023]. | IQR |
| Root Mean Square | The square root of the average of the squared amplitude values of the signal Doulah et al. [2014]. | RMS |
| Signal Magnitude Area | The normalized integral of the absolute value of the signal's amplitude Liu et al. [2014]. | SMA |
| Energy | The sum of the squared amplitude values of the signal Korda et al. [2015]. | ER |
| Entropy | The randomness or unpredictability in the signal values Wunderlich and Gramann [2021]. | EN |
| Zero Crossing Rate | The frequency at which the signal's amplitude crosses zero, indicating sign changes over a period Wunderlich and Gramann [2021]. | ZCR |
| Area Under Curve | The area under the curve of the signal's amplitude versus time Aungsakul et al. [2012]. | AUC |
| Spectral Centroid | The center of mass of the signal's power spectrum Starzacher and Rinner [2009]. | SC |
| Spectral Bandwidth | The range of frequencies over which the signal's power is distributed, measuring the width of the signal's power spectrum Starzacher and Rinner [2009]. | SB |
| Lyapunov Exponents | A measure of how quickly nearby trajectories in the signal's phase space diverge or converge, indicating sensitivity to initial conditions and chaotic behavior Kim and Park [2001]. | LE |
| Detrended Fluctuation Analysis | A method for measuring long-range correlations in non-stationary time series by detrending and analyzing fluctuations at various scales Li et al. [2015]. | DFA |
| Teager Kaiser Energy Operator | A nonlinear measure of instantaneous signal energy, emphasizing rapid amplitude variations Zeng et al. [2021]. | TK |

Table 2: Results of Indices Generated by the Proposed Algorithm

| Score Name | R2 Score | MAE | MSE | RMSE | MAPE | $\rho$ |
|---|---|---|---|---|---|---|
| **Navigation and Orientation** | 0.72 | 4.16 | 25.58 | 5.057 | 7.76 | -0.81 |
| **Spatial Anxiety** | 0.51 | 4.97 | 41.81 | 6.466 | 20.21 | -0.63 |
| **Distance Estimation** | 0.33 | 2.86 | 13.93 | 3.73 | 37.01 | -0.56 |
| **Landmark Recognition** | 0.50 | 0.78 | 0.86 | 0.93 | 12.99 | -0.63 |
| **Path Route** | 0.52 | 0.43 | 0.28 | 0.53 | 20.25 | -0.79 |
| **Path Survey** | 0.41 | 0.68 | 0.80 | 0.89 | 34.26 | -0.63 |
| **Location Allocentric** | 0.08 | 0.80 | 1.04 | 1.02 | 39.37 | -0.31 |

Table 3: Spearman Correlation of Statistical Features with Eye Movements

| Blink Count | Fixation Count | Fixation Duration | Saccade Count | Saccade Duration |
|---|---|---|---|---|
| $MD_v$ (-0.81) * | $DFA_h$ (-0.37) | $DFA_h$ (0.36) | $TK_h$ (0.42) | $SK_h$ (-0.51) * |
| $TK_v$ (0.80) | $ZCR_h$ (0.30) | $SK_v$ (0.36) | $TK_v$ (-0.37) | $MO_h$ (-0.46) * |
| $ER_v$ (0.77) * | $SK_v$ (-0.28) | $SMA_v$ (-0.36) | $MA_v$ (-0.37) | $MI_h$ (-0.46) * |
| $RMS_v$ (0.77) * | $MA_v$ (-0.24) | $AUC_v$ (0.32) | $AUC_v$ (-0.35) | $AUC_v$ (-0.42) |
| $SMA_v$ (0.73) | $TK_h$ (0.24) | $TK_h$ (-0.31) | $DFA_h$ (-0.33) | $SB_v$ (-0.41) * |
| $MO_v$ (-0.55) | $KU_h$ (-0.24) * | $MO_v$ (0.30) | $SK_v$ (-0.27) | $MO_v$ (-0.38) |
| $ZCR_v$ (0.51) | $ZCR_h$ (-0.28) | $ZCR_v$ (0.19) | $RA_v$ (-0.23) * | $EN_v$ (0.33) * |